\documentstyle[12pt,epsf]{article}
\begin{document}
\baselineskip 18pt

\newcommand{\ba}{\begin{array}}
\newcommand{\ea}{\end{array}}
\newcommand{\bd}{\begin{displaymath}}
\newcommand{\ed}{\end{displaymath}}
\newcommand{\be}{\begin{equation}}
\newcommand{\ee}{\end{equation}}
\newcommand{\bea}{\begin{eqnarray}}
\newcommand{\eea}{\end{eqnarray}}
\def\bra{\langle}
\def\ket{\rangle}
\def\a{\alpha}
\def\b{\beta}
\def\g{\gamma}
\def\d{\delta}
\def\e{\epsilon}
\def\ve{\varepsilon}
\def\l{\lambda}
\def\m{\mu}
\def\n{\nu}
\def\G{\Gamma}
\def\D{\Delta}
\def\L{\Lambda}
\def\s{\sigma}
\def\p{\pi}

\vskip .6cm

\begin{center}
{\Large Baryogenesis in a supersymmetric model without $R-$parity}\\ 
\vskip .5in
{\large Rathin Adhikari\footnote{
e-mail address : rathin@prl.ernet.in} and Utpal Sarkar}\\[.3in]

{\large Theory Group }\\
{\large Physical Research Laboratory} \\
{\large Ahmedabad - 380009, India.}

\end{center}

\vskip .75in
\begin{abstract}
\baselineskip 18pt

We propose a simple scenario for  baryogenesis in  supersymmetric
models where  baryon  number is broken  alongwith  R-parity.  The
lightest  supersymmetric  particle  (neutralino)  decays to three
quarks and $CP-$ violation  comes from  interference  of tree and
one loop box  diagrams.  The  bounds on the  $R-$parity  breaking
couplings from the out-of-equilibrium  condition are considerably
relaxed in this scenario.

\end{abstract}

\newpage
\baselineskip 18pt

The  observable  universe  contains  more matter than  antimatter
although the ratio is very small for baryons.  This  asymmetry of
the universe \cite{olive} can be generated at a very high energy,
but in most  likelyhood  it will be washed  out at a later  stage
\cite{hooft}.  So a great deal of interest  started in  scenarios
where the baryon  asymmetry is generated  during the  electroweak
phase   transition   \cite{ewbar,shapos}   by  making  the  phase
transition to be a weakly first order.  However the condition for
survival  of the  generated  baryon  asymmetry  after  the  phase
transition  gives a strong bound on the mass of the higgs doublet
\cite{higgsbound}, which rule out this scenario.  This  motivates
for alternative scenarios for baryogenesis [6-10].

In this  article  we  propose  a simple  supersymmetric  model to
generate baryon  asymmetry of the universe at low energy.  Baryon
number   violation   arises  from  R-parity   breaking.  R-parity
violating couplings, which also violate lepton number, is assumed
to be small (or even zero if lepton  number  is  conserved).  The
lightest  supersymmetric  particle  (LSP),  which  is  one of the
neutralinos $\chi_1^0$, is now unstable since R-parity is broken.
There exist new box-type  diagrams  for the decay of  $\chi_1^0$,
which  interferes  with the tree  level  diagram  to give  enough
$CP-$violation,  which  generates  the  baryon  asymmetry  of the
universe.  The mass of  $\chi_1^0$ is much less than the sfermion
masses and as a result when  $\chi_1^0$  decays the sfermions has
already  decayed  away, so it does not  matter  if the  sfermions
decay in  equilibrium  and  have  erased  the  primordial  baryon
asymmetry.  This  relaxes  the  bounds on the  R-parity  breaking
B-violating couplings considerably.

In supersymmetric  models, R-parity was introduced as a matter of
convenience  to prevent  fast proton  decay.  It is now  realised
that the proton lifetime can be made consistent  with  experiment
without invoking R-parity  symmetry.  If we don't impose R-parity
in the model,  then the  minimal  supersymmetric  standard  model
allows  the  following  $B$  and  $L$  violating   terms  in  the
superpotenial
\begin{equation}
W = \lambda_{ijk} L^i L^j {\left(E^k \right)}^c +
\lambda^{\prime}_{ijk} L^i Q^j {\left( D^k \right) }^c +
\lambda^{\prime \prime}_{ijk}{\left( U^i \right) }^c{\left( D^j
\right) }^c {\left( D^k \right) }^c  
\end{equation}
Here $L$ and $Q$ are the  lepton and quark  doublet  superfields;
$E^c$ is the lepton  singlet  superfield  and $U^c$ and $D^c$ are
the quark singlet  superfields  and $i, j, k$ are the  generation
indices.  In the above the first  two  terms  are  lepton  number
violating  while the third term violates  baryon number.  For the
stability   of  the  proton,  we  assume  that   $\l_{ijk}$   and
$\l^{\prime}_{ijk}$  are  extremely  small or even  zero (if some
symmetry   like  lepton   number  is   present).  The   couplings
$\l^{\prime  \prime}_{ijk}$ can now be considerably  large.  They
are antisymmetric in the last two indices and in general complex.

We consider one of the neutralinos  ($\chi_1^0$)  as the lightest
supersymmetric  particle  having  mass of the order of 100 -- 200
GeV and the sfermions have higher mass of the order of 250 GeV to
a few TeV.  So before the universe cools down to the  temperature
of the  electroweak  symmetry  breaking scale the sfermions  have
already  decayed  away.  Since  R-parity  is broken,  some of the
sfermions will have baryon number violating decay channels, which
may  even  wash  out  the  primordial  baryon  asymmetry  of  the
universe.

Near the  electroweak  scale the  neutralinos  $\chi_1^0$ are the
only  superparticles  left  to be  dacayed.  If  R-parity  is not
broken, these  particles are stable.  However, since  R-parity is
now broken, the neutralino can also decay to ordinary  quarks and
leptons.  In these  decays at least one of the vertex  should not
conserve  R-parity and hence baryon number is broken.  In figures
1(a) and 2(a) the tree level diagrams for the processes $\chi_1^0
\to u_{iL}  {d_{jR}} { d_{kR}}$  and $u_{iR}  {d_{jL}} { d_{kR}}$
are given, while in figures 1(b) and 2(b) we present the new type
of one loop box diagrams.  The interference of the tree level and
the  one  loop  diagrams  give  rise  to  the CP  violation.  The
advantage  of the  box  diagrams  is  that  two  of the  vertices
contains  couplings  of the higgs and hence  elements  of the CKM
matrix  contributes to the baryon  asymmetry.  This gives a large
enhancement  and hence we can get a large baryon  asymmetry  even
for considerably smaller R-parity breaking couplings.

As  the  $B$-violating   couplings  are  antisymmetric  with  the
interchange  of the d-type quark indices, the indices $j$ and $k$
are always  different  for a particular  type of decay mode where
$i$, $j$ and $k$ are fixed.  Thus with the interchange of $j$ and  
$k$ there will be an extra diagram at the tree level corresponding
to figure 2a  and  two other box diagrams corresponding to figures 
1b and 2b.

The thermally   averaged decay width for such baryon number
violating decays   are  given by  
\bea  
\bra \G ( \chi_1^0 \to   u_{i} d_{j} d_{k})  
\ket  &\approx& {g^2 
{\mid \l_{ijk}^{\prime \prime}  \mid}^2  m_{\chi_1^0}^5 
\over  10^{4} \pi^4 {\left( m_{\tilde q}^2 + T_{0}^2   \right) }^2
}  A^2 
\eea
where the factor $A$ for the couplings $\bar{u}_R \tilde{u}_R {\chi_1^0}$
and $\bar{d}_R \tilde{d}_R {\chi_1^0}$ are 
$$ A = {\left[\pm N_{12} /2 + {\tan \theta_w/6 }\; N_{11} \right]}$$
respectively. We have followed the notations of Haber and Kane
\cite{Haber} for the MSSM type squark-quark-neutralino interaction. 

The   decay   of  the   neutralino   should   now   satisfy   the
out-of-equilibrium  condition  near  $T=m_{\chi_1^0}$,  which  is
$\bra \G (  \chi_1^0  \to u_{i}  d_{j}  d_{k} ) \ket < 1.7 {\sqrt
g_*} (T^2/M_P) $.  This can be satisfied with $A \sim 10^{-2}$ to
$10^{-3}$,  which is possible for a wide range of MSSM parameters
for neutralino  mass ranging from about 100 to 200 GeV with $\mid
\mu \mid$  about 200 to 1000 GeV and $\tan  \beta$  from 2 to 12,
and with $\l^{\prime \prime} < 10^{-1}$ to $10^{-2} $.

For the  decay  modes  of  figures  1(a)  and  2(a),  we  require
$\bar{u}_L  \tilde{u}_R  {\chi_1^0}$  and $\bar{d}_L  \tilde{d}_R
{\chi_1^0}$ couplings, which comes from the higgsino component of
the  neutralino.  In ref  \cite{Haber}  these  couplings  are not
present since they assumed $m_q = 0$.  In general, the neutralino
will have a higgsino component, coming from the mixing at the one
loop level with the  internal  lines as quarks and  squarks.  The
mass  insertion  for the quarks will allow a change in  helicity.
The  dominant  contribution  comes from the top quark in the loop
and hence the suppression  factor is  proportional to $m_t$.  The
coupling of the  neutralino  $\bar{q}_L  \tilde{q}_R  {\chi_1^0}$
will  thus be  suppressed  by a factor of \be F = {1 \over 2 \pi}
{m_t \over  m_{\tilde{q}}}  \ee compared to the  couplings $A$ of
$\bar{q}_R  \tilde{q}_R  {\chi_1^0}$.  Instead of considering the
coupling  $\bar{q}_L  \tilde{q}_R  {\chi_1^0}$  of  the  higgsino
component of the neutralino with the suppression factor $F$ it is
also  possible  to  take  the  neutralino   coupling   $\bar{q}_R
\tilde{q}_R {\chi_1^0}$ and then change the helicity of the quark
by a mass  insertion.  But that will then introduce a suppression
by  $m_q/m_{\tilde{q}}$  and hence we do not consider  this later
possibility.

While  calculating  the  baryon  asymmetry  in the  decay  of the
neutralino  $\chi_1^0$,  one has to take into  consideration  the
fact that some of the  neutralinos  may  thermalize  before  they
decay  through the $ \chi_1^0 \; q  \rightarrow  \chi_1^0  \; q $
scattering.  This will introduce an additional suppression factor
\be
S =  {5 \over 4 \pi^2 g^2} { {\mid  \l^{\prime \prime}_{ijk}  \mid}^2
\over  A^2  }  {\left( { m_{\chi_1^0}  \over  T_0 } \right)}^5  .
\ee

Taking the  suppression  factors $S$ and $F$ into  account we can
now  calculate the baryon  asymmetry $\e $ generated  through the
interference of the various possible tree level diagrams [figures
1(a) and 2(a)] with the box diagrams [figures 1(b) and 2(b)]
\begin{equation}
\e  = { 5  \over 2 \pi^3 }{ (m_{d_j} m_{d_l} + m_{u_i} m_{u_n})
\over m_W^2}{ m_t^2 \over m_{\tilde q}^2}
{m_{\chi_1^0}^5 \over  T_0^5 }   
{ \tan^2 \beta \over  A^2} 
[ {\rm Im} \left( {\l^{\prime \prime}_{ijk}}^*  \; \l^{\prime
\prime}_{nlk}  V_{n^\ast j} V_{i^\ast l}^\ast  \right) ]
 I({m_{\chi_1^0} \over  m_{\tilde{q}}})  \label{eps}
\end{equation}
where $I({m_{\chi_1^0} / m_{\tilde{q}}})$ comes from the  
absorptive  part of the loop integral. For $m_{\chi_1^0} \ll m_{\tilde{q}}$
it may be approximated as,
\be
I({m_{\chi_1^0} \over  m_{\tilde{q}}}) =   
{ 1 \over 32 \pi } \; {m_{\chi_1^0} \over  m_{\tilde{q}}} .
\ee
$V_{a^* b}$ is the CKM matrix,  which  enters from the
higgs  coupling.  It  is  clear  from  the  expression  that  the
imaginary part of the product of the couplings is invariant under
rephasing  of all the  quark  phases.  We have  assumed  that the
neutralino   couplings  are  diagonal   since  the  off  diagonal
couplings will be further  suppressed.  It can be noticed that it
is possible to choose a phase  convention, so that the $\l$'s are
real and hence the  $CP-$violating  phase comes entirely from the
CKM  matrix.  In this  basis,  there  will not be any  constraint
coming  from the  electron  dipole  moment of the  neutron on the
complex  part of the  couplings  of $\l$.  This  will
reduce  one more  uncertain  parameter  in the  problem.

In equation (\ref{eps}) it is possible to choose the parameters
so that the amount of asymmetry generated in this scenario is
large enough. As a representative set of values we consider 
$T_0 \approx m_{\chi_1^0} $. We consider $
\tan \beta \sim  3$; mass of the neutralino of the order
of  200 GeV and the mass of squarks are in the range of 
a few TeV. Since top quark in the decay product will be 
phase space suppressed, we consider the decay products to be
$c_R b_L d_L$. The internal lines have top and strange quarks
and hence the two CKM matrix elements are $V_{tb}$ and $V_{cs}$.
Then for $A^2 \sim  10^{-5} $ we get $\e \sim 10^{-2}\times  
\l^{\prime  \prime}_{ijk} \; \l^{\prime \prime }_{nlk} $. In
this case large enhancement comes from the mass of the top quarks
and diagonal CKM elements. For several other choices of parameters
also we can have a large $\e$ in this scenario.

We  shall  now  discuss  the  constraints  on  the  $B-$violating
R-parity  breaking coupling constant  $\l^{\prime  \prime }$.  In
general  one  requires  that  all  $B-$violating  decays  of  the
sfermions  should be slow  enough  so as not to erase the  baryon
asymmetry before the electroweak phase transition  \cite{salati}.
However, in the  present  model these  constraints  are no longer
valid since we generate baryon asymmetry at the electroweak scale
when all the sfermions have already decayed away.

The constraint  coming from the out of equilibrium condition for
the interaction $ u_i d_j \rightarrow  {\bar {d_k}} \chi_1^0 $
is usually very severe \cite{salati}
\be
\l^{\prime \prime}_{ijk}    <        3       \times
10^{-6  } {\left( {\tilde m}  \over 1 TeV \right)}^{  1/2}
{\left[ {\sqrt 2} g    {    \tan \theta_w  \over 3}  N_{12}      
 \right] }^{-1} .
\ee                
For the mass of neutralino of the order
of 100 to 200 GeV this  bound  of $\l^{\prime \prime}_{ijk}$
is of the order of $10^{-3}  $ for
some region of the parameter space  independent of the
generation indices. However, in the present scenario the CP violating
part of this diagram will also contribute to the generation of baryon 
asymmetry. This means that even if this constraint is not satisfied,
the suppression will be linear in $\l^{\prime \prime}_{ijk}$ and this
constraint is also relaxed.

Another source of constraint on $\l^{\prime \prime}_{ijk}$ comes
from the non-observation of $n {\bar n} $ oscillations. 
The  earlier bound \cite{zwirner,barbi}  was  
$
\l^{\prime \prime}_{11k}  \ll    2 \times 10^{-7} {\left[ { \tilde m
/ 100 GeV}  \right]}^{5/2} .
$
However these constraints are highly model dependent and
may be evaded \cite{Drei,Sher} and the upper bound may be of
the order of $10^{-3}$  or higher depending on the choice
of SUSY parameters. The coupling 
$\l^{\prime \prime}_{113}$ is almost free of any constraint, when a
suppression factor coming from the flavor changing neutral current
is included \cite{Sher}. For $ \l^{\prime \prime}_{112}  $ 
the upper bound  may vary from $10^{-4} $ to  $10^{-  1 } $
depending on the  stop mass  from 100 Gev to    500 GeV \cite{Sher}. 
In our case the upper bound on these couplings will be still higher.  

The product of two $B $ violating  couplings are constrained from
the rare  two body  non-leptonic  decays  of $B $ and $D $ mesons
\cite{Sher}.  However  those  constraints  are much  weaker.  For
higher  squark  mass in the TeV range most  stringent  constraint
essentially   comes  from  the  out  of  equilibrium   conditions
\cite{Sher,Keung} for most of the $B-$violating  couplings, which
is of the order of $10^{-2}$ to $10^{-3}$.  But these constraints
are not applicable in the present scenario as discussed  earlier.

Satisfying  all the  constraints  it is  thus  possible  to  have
$\l^{\prime \prime }$ as high as $10^-1$.  However for the choice
of parameters we considered, we require only $\l^{\prime \prime }
\sim  10^{-3}$  to get $\e$ as high as  $10^{-8}$.  This gives us
enough  freedom to relax some other  constraints  and still  have
enough baryon  asymmetry.  In this scenario this baryon asymmetry
will be a $(B-L)$ asymmetry, and hence this is not washed out due
to  sphaleron  transition.  On the other  hand  since the  baryon
asymmetry is not related to the lepton asymmetry of the universe,
all  constraints  on the  lepton  number  violating  interactions
arising due to the Majorana masses of the neutrinos are no longer
valid.

To  summarize,  we  presented  a simple  supersymmetric  model of
baryogenesis,   where   baryon   number  is  violated   alongwith
$R-$parity.  $CP-$violation  comes  from an  interference  of the
tree level and the one loop box-type  diagrams,  such that two of
the elements  enters from the CKM matrix.  The constraints on the
$B-$violating  ($R-$parity  violating) couplings are considerably
relaxed in this scenario.

{\bf  Acknowledgement}  One of us (US)  acknowledges a fellowship
from the Alexander von Humboldt  Foundation and hospitality  from
the Institut  f\"{u}r  Physik, Univ Dortmund,  where part of this
work was done.

\newpage

\begin{figure}[htb]
\mbox{}
\vskip 6.0in\relax\noindent\hskip -0in\relax
\includegraphics{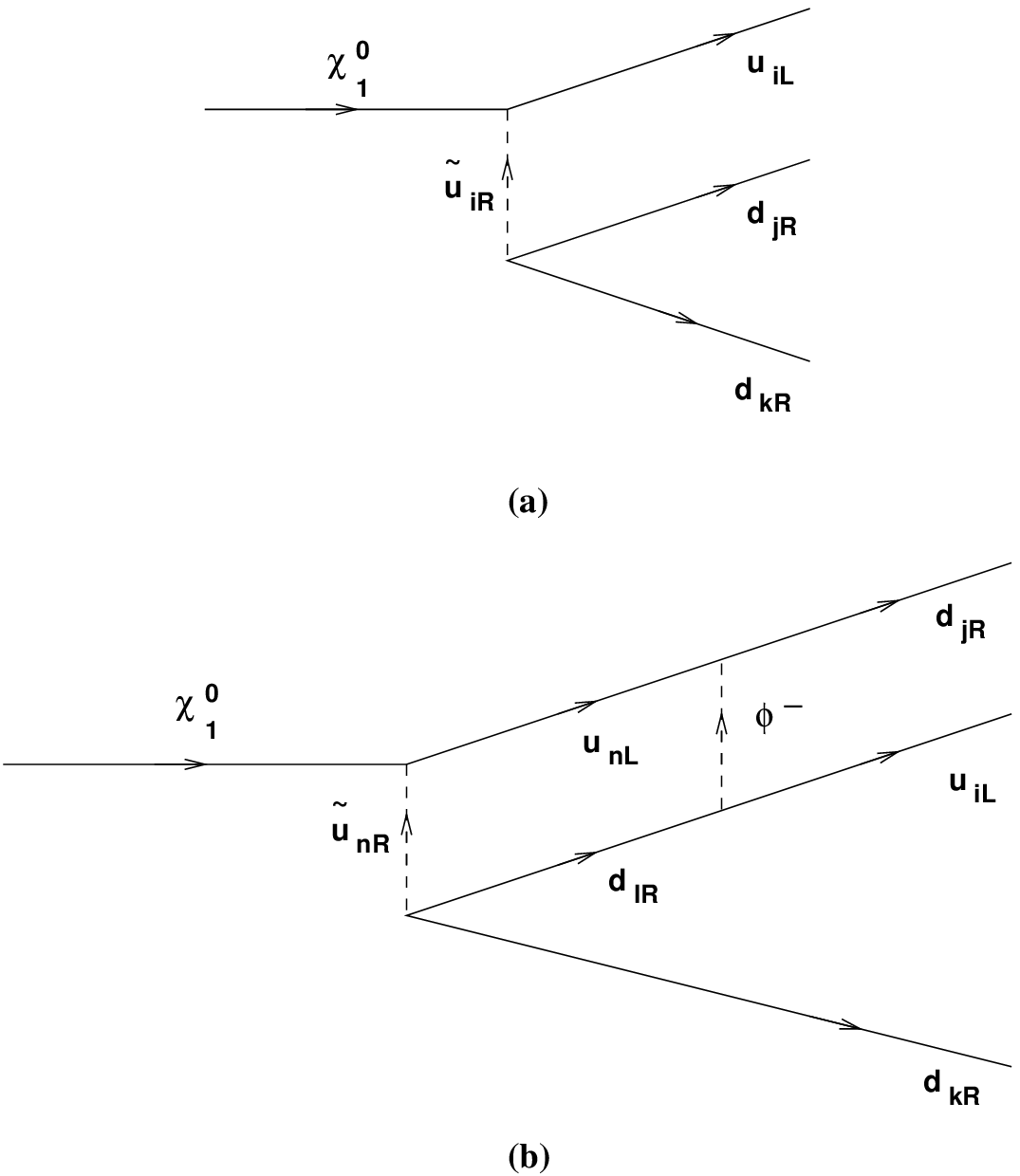} \vskip .25in
\caption{ Tree level and box diagrams for the decay $
\chi_1^0  \rightarrow   u_{iL} d_{jR} d_{kR} $.}
\end{figure}

\begin{figure}[htb]
\mbox{}
\vskip 6.0in\relax\noindent\hskip -0in\relax
\includegraphics{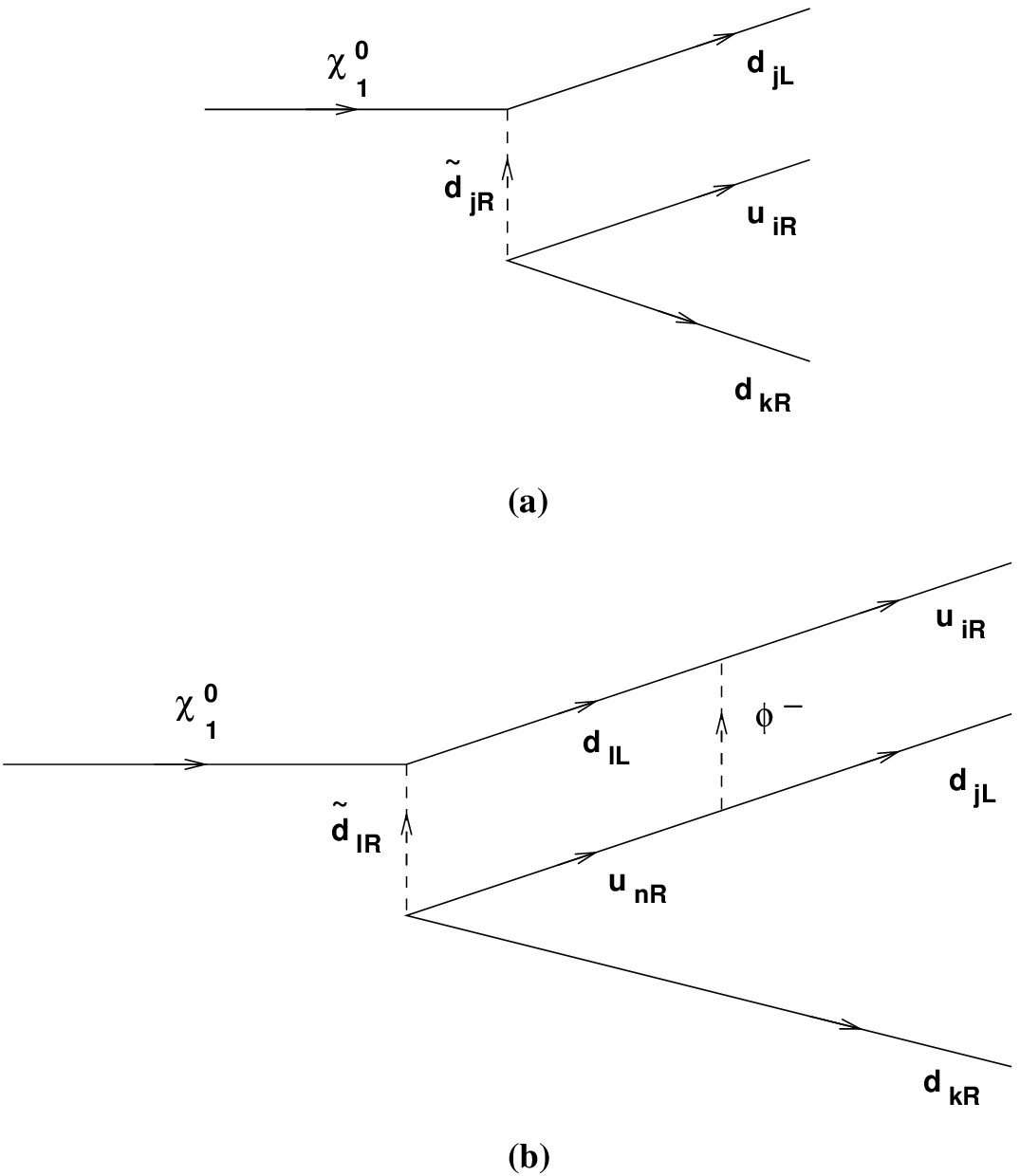} \vskip .25in
\caption{ Tree level and box diagrams for the decay $
\chi_1^0  \rightarrow   d_{iL} u_{jR} d_{kR} $.}
\end{figure}

\end{document}